\documentclass[prd,preprint,aps,tighten,nofootinbib,11pt]{revtex4}

\usepackage{amsmath,graphicx,color,epsfig}

\begin{document}

\preprint{DESY 11-228}

\title{Factorization of Heavy-to-Light Baryonic Transitions in SCET}

\author{ Wei Wang~\footnote{Email:wei.wang@desy.de}}
\affiliation{Deutsches Elektronen-Synchrotron DESY, D-22607 Hamburg, Germany}

\date{\today}

\begin{abstract}
In the framework of the soft-collinear effective theory, 
we demonstrate that  the leading-power heavy-to-light baryonic form factors  at large recoil  obey the heavy quark and large energy symmetries.  Symmetry breaking effects   are suppressed by  ${\Lambda}/m_{b}$ or  $\Lambda/E$, where $\Lambda$ is the hadronic scale,  $m_b$ is the $b$ quark mass and  $E\sim m_b$ is the energy of light baryon in the final state. At leading order, the leading power baryonic form factor $\xi_{\Lambda,p}(E)$, in which two hard-collinear gluons are exchanged in the baryon constituents,  can factorize into the soft and collinear matrix elements convoluted with a  hard-kernel of order $\alpha_s^2$.  Including the  energy release dependence, we derive  the  scaling law   $\xi_{\Lambda,p}E)\sim \Lambda^2 /E^2$. We also find that this form factor $\xi_{\Lambda}(E)$  is numerically smaller than the form factor   governed by soft processes, although the latter is formally power-suppressed. 

\vspace{1.cm}
\textbf{Keywords:} Heavy quark physics,  QCD, b-physics
\end{abstract}

\maketitle


\section{ Introduction}

Precision test of the unitarity of the CKM matrix, allowing us to explore the SM description of the CP violation and reveal  any physics beyond the SM,  greatly depends on our knowledge of the nonperturbative matrix elements. Fortunately the calculation of the amplitudes  of   bottom meson decays is under  control as the amplitudes can be  expanded in terms of   small ratios justified
by both the large $b$-quark mass, and a large energy release in the decay.  With this expansion, a number of theoretical predictions on different observables in various channels  are    found in global agreement with experimental measurements (see Ref.~\cite{arXiv:0801.1833} for a review). 

Decay processes of heavy baryons consisting of a bottom quark provide complementary information with the $B$ meson and thereby are receiving growing attentions on both  experimental and theoretical sides. Semileptonic decays, such as $\Lambda_b\to p l\bar\nu$, are simplest exclusive baryonic decays and  governed  by heavy-to-light form factors.  In this retrospect,  apart from the theoretical analysis    based on the heavy quark effective theory~\cite{UTPT-90-03,Mannel:1990vg,hep-ph/9701399}, the simplification of   baryonic form factors in the large energy limit is  exploited~\cite{Feldmann:2011xf,Mannel:2011xg} (see Ref.~\cite{Hiller:2001zj,hep-ph/0702191} for an earlier discussion), applying  the method developed in the mesonic case~\cite{hep-ph/9812358,HUTP-90-A071}.  In the $\Lambda_b\to \Lambda$ transition,  only one    form factor is nonzero after the reduction and this universal  function (soft form factor) is also calculated within the  light-cone QCD sum rules in conjunction with  the effective field theory~\cite{Feldmann:2011xf}.

Soft-collinear effective theory (SCET)~\cite{Bauer:2000yr,Bauer:2001cu,hep-ph/0109045,Beneke:2002ph,hep-ph/0211069}  is a powerful tool to describe  processes with   particles having energy much larger than their mass.  The heavy-to-light decay of heavy baryons, for instance $\Lambda_b\to  p\l\bar\nu$,  is of this type. SCET makes use of the expansion in small ratios, in this case, $\lambda= \sqrt {\Lambda/m_b}$ with $\Lambda$ as the hadronic scale and $m_b$ as the $b$ quark mass.  One of the most important features of SCET is that the interaction between the soft and collinear sectors is   taken into account, overcoming the shortcomings in the  large energy effective theory~\cite{hep-ph/9812358,HUTP-90-A071}.  
Therefore in SCET  not only the reduction of the leading-power form factors
 is formulated on  the QCD  basis, but  also the symmetry-breaking corrections can be systematically explored~\cite{hep-ph/0211018,hep-ph/0508250}.

In this work, we will analyze the baryonic form factors in SCET and follow the techniques developed in the  $B\to \pi$ form factor which  takes the following factorization form at the leading power~\cite{hep-ph/0008255,hep-ph/0211069,Beneke:2003pa} 
\begin{eqnarray}
 F_i^{B\to \pi} (E)= C_i \xi_\pi (E)+ \int d\tau C_i' (E,\tau) \Xi_{\pi}(\tau, E).
\end{eqnarray}
Here  $E$ is the energy of the final hadron and $C_i$ and $C_i'$ are the short-distance coefficients obtained by matching from QCD onto the effective field theory. The one-loop expressions for these  coefficients   can be found  in Refs.~\cite{hep-ph/0008255,Bauer:2000yr,hep-ph/0109045,Beneke:2003pa,hep-ph/0402241,hep-ph/0408344}. 
In what follows we will adopt the ansatz that the final light particle is composed of collinear objects and thus  hard-collinear gluon exchange is required to turn the soft spectators into energetic ones. 
 In such picture,  to the end  we will show that  the matrix elements  parametrizing   form factors, in the example of $\Lambda_b\to \Lambda$,   are formally simple
\begin{eqnarray}
 \langle\Lambda (p')|\bar s\Gamma b|\Lambda_b(p)\rangle = C_i \xi_\Lambda (E)\bar u_{\Lambda}(p')\Gamma u_{\Lambda_b}(p) + {\cal O}(\lambda^2 \xi_{\Lambda}),\label{eq:formfactorbaryonSCET}
\end{eqnarray}
in which the spin indices are suppressed.  For  contributions dominated by soft processes which are not suppressed by $\alpha_s$, please see Refs.~\cite{Feldmann:2011xf,Mannel:2011xg}.

The remainder of this work is organized as follows. In Sec.~\ref{SCET:analysis}, we will present the form of the leading power and next-to-leading power heavy-to-light currents  in SCET after integrating out the hard modes,  and following  Ref.~\cite{Beneke:2003pa} discuss their representations in the effective theory containing soft and collinear modes.  In Sec.~\ref{eq:QCDanalysis}, the transition form factors are directly calculated in QCD, and we   show the correspondence with the SCET effective operators. Several implications from our analysis are given in Sec.~\ref{sec:discussions}, and a summary of our findings is presented in Sec.~\ref{sec:conclusion}.

\section{SCET analysis} \label{SCET:analysis}

We use the position-space representation of SCET and closely follow the notations in Refs.~\cite{Beneke:2002ph,Beneke:2003pa}. We work in the b-baryon rest frame and use the light-cone coordinate, in which  a momentum $p$ is decomposed as
\begin{eqnarray}
 p^{\mu} = (n_+p)\frac{n_-^\mu}{2} +p_{\perp}^\mu + (n_-p) \frac{n_+^\mu}{2} 
\end{eqnarray}
where $n_\pm$ are two light-like vectors: $n_+^2=n_-^2=0$ and $n_+\cdot n_-=2$.  The reference directions $n_\pm$ are chosen such that the energetic massless external lines in the recoiling system have $n_+p$ of order $m_{b}$, while the magnitude of $n_-p$ is small.  
This type of  momenta  is  collinear: 
$ p_c = (n_+p, p_\perp, n_-p) \sim (1, \lambda^2, \lambda^4)$. The slowly-moving degrees of freedom in the heavy   baryon have soft momenta
$
 q_s\sim (\lambda^2, \lambda^2, \lambda^2).$ For the heavy $b$ quark, the statement of ``soft" refers to  the residual momentum after removing  the large component which becomes a label of heavy quark.  The hard-collinear mode, with ${\cal O}( m_b\Lambda)$ virtuality,  arises  from the interaction between soft and collinear sector: $
 p_{hc} \sim (1, \lambda, \lambda^2) $. 

Power scalings of quark and gluon fields are determined  by the configuration of their momenta.  For the quark fields, we have
\begin{eqnarray} 
 \xi_c=\frac{n\!\!\!\slash_-n\!\!\!\slash_+}{4} \psi_c\sim \lambda^2,  \;\;
 \xi_{hc}=\frac{n\!\!\!\slash_-n\!\!\!\slash_+}{4} \psi_{hc}\sim \lambda, \nonumber \\
 q_s\sim \lambda^3,\;\; h_v = \frac{1+v\!\!\!\slash}{2} Q_v \sim \lambda^3. 
\end{eqnarray} 
Here $v$ is the velocity of the heavy quark.  $\xi_{c,hc}$ and $h_v$ are   large components of the collinear, hard-collinear and heavy quark fields, respectively. 
Small components of the heavy quark field, $H_v$,   and  collinear quarks, $\eta_{hc}$ and $\eta_c$, can be integrated out at tree level by solving the equation of motion. Scalings of  gluon fields have a similar behavior with  their momenta
\begin{eqnarray}
 n_{+}A_c\sim 1,\;\
 n_{-}A_c\sim \lambda^4,\;\
 A_{\perp c}\sim \lambda^2 ,\;\ A_s\sim \lambda^2, \nonumber\\
 n_{+}A_{hc}\sim 1,\;\
 n_{-}A_{hc}\sim \lambda^2,\;\
 A_{\perp hc}\sim \lambda. 
\end{eqnarray}
From the relativistic normalization condition, we find that  the   
baryonic states in the effective theory, taking the $\Lambda_b$ and $\Lambda$ as an example,  have the scaling
\begin{eqnarray}
 |\Lambda_b\rangle \sim \lambda^{-3}, \;\; |\Lambda \rangle \sim \lambda^{-2},
\end{eqnarray}
where we did not specify the differences with the states in QCD. 
Presumably these differences may introduce more power corrections, but  they are left out here, since  the leading-power behavior  is unlikely to change. 
Decay constants of baryons defined via~\cite{arXiv:0804.2424,arXiv:0811.1812}
\begin{eqnarray}
 \epsilon^{ijk} \langle 0|( u^i C\gamma_5 d^j) h_v^k |\Lambda_b\rangle = f_{\Lambda_b}^{(1)} u_{\Lambda_b}, \;\; 
 \epsilon^{ijk} \langle 0|( u^i C\gamma_5 v\!\!\!\slash d^j) h_v^k |\Lambda_b\rangle = f_{\Lambda_b}^{(2)} u_{\Lambda_b},  \nonumber\\
 \epsilon^{ijk} \langle 0|( u^i C\gamma_5 \frac{n\!\!\!\slash_+}{2}  d^j) \frac{n\!\!\!\slash_+}{2}  s^k |\Lambda(p')\rangle = f_{\Lambda} \frac{n_+p'}{2} \frac{n\!\!\!\slash_+}{2}   u_{\Lambda},   \nonumber
\end{eqnarray}
scale as $f_{\Lambda_b} \sim \lambda^6$  and $f_{\Lambda}\sim \lambda^4$ with $f_{\Lambda_b}$ denoting both $f_{\Lambda_b}^{(1)}$ and $f_{\Lambda_b}^{(2)}$.

In SCET, integration of the fluctuations with large virtualities proceeds in two-steps~\cite{Beneke:2003pa,hep-ph/0211069}.  In the first step,  hard scales, caused by the interaction between the collinear sector and heavy quark, and   between two or more collinear sectors with different directions, are integrated out and thereby QCD is matched onto   an intermediate effective theory, called   SCET$_{I}$.   In this effective theory   gauge invariant operators are  built out of  fields of hard-collinear quarks or soft  gluons and quarks.  The leading-power and next-to-leading power terms having non-zero matrix elements between the baryonic transition are constructed as
\begin{eqnarray}
O^{A}_j(s)= (\bar \xi_{hc}  W_{hc} )_{s} \Gamma_j  Y_s^\dagger h_v,\nonumber\\
 O^{B}_j(s_1,s_2)=   (\bar \xi_{hc} W_{hc})_{s_1} (W^\dagger _{hc}   iD_{\perp   \mu }  W_{hc})_{s_2}  \Gamma_j'  Y_s^\dagger h_v,\nonumber\\
 O^{C}_j(s_1,s_2,s_3)= (\bar \xi_{hc}  W_{hc})_{s_1}   (W^\dagger _{hc}    iD_{\perp  \mu_1}  W_{hc})_{s_2} (W^\dagger _{hc}   iD_{\perp  \mu_2 }  W_{hc}) _{s_3} \Gamma_j' Y_s^\dagger h_v,
\nonumber\\
 O^{D}_j(s_1,s_2)= (\bar \xi_{hc} W_{hc})_{s_1} (W_{hc}^\dagger in_-D W_{hc} -in_-D_s)_{s_2} \Gamma_j'   Y_s^\dagger  h_v,\nonumber\\
 O^{E}_j(s,t)= (\bar \xi_{hc}  W_{hc})_{s}  (iD_s^\mu)_{t} \Gamma_j' Y_s^\dagger h_v,\label{eq:SCETIoperator}
\end{eqnarray} 
where the hard-collinear field with the subscript $s$ is evaluated at $x+s n_+$, while the soft field with the subscript $t$ is evaluated at $x+t n_-$, with $x$ being the space coordinate from the QCD current.    $\Gamma_{j}'$ is one of the following gamma matrices 
\begin{eqnarray}
 \Gamma_j'=(1,\gamma_5, \gamma_{\perp}, \gamma_{\perp}\gamma_5).  \label{eq:Lorentz-structures}
\end{eqnarray} 
The $W_{hc}$ and $Y_s$ are   hard-collinear and soft Wilson lines, respectively~\cite{Beneke:2003pa,hep-ph/0211069}.

Integration of the hard-collinear   mode will result in the final SCET, named as SCET$_{II}$ for convenience.  In SCET$_{I}$,  the generic power scalings  of  the operators in Eq.~\eqref{eq:SCETIoperator}  are
\begin{eqnarray}
 O^A\sim \lambda^4, \;\; Q^{B} \sim \lambda^5, O^{C,D,E}\sim \lambda^6.
\end{eqnarray}
But none of them  have the  right quantum numbers with   baryons in the initial and  final state.  
 Thus the  matching  of these operators from SCET$_{I}$ onto SCET$_{II}$ will induce additional power suppressions and one of our goals is to count these suppressions. 
To the end, 
we will demonstrate  that the contribution from the $O^A$ operator starts at the ${\cal O}(\lambda^9)$, while the other types of operators have the power $\lambda^{11}$.


\subsection{General analysis in SCET$_{II}$}

To represent the quantum numbers of the  $\Lambda_b$ and $\Lambda$ baryon, at least the fields $q_s q_s h_v$ and three collinear quark fields are needed.   In the light-cone gauge 
a most  general form of an operator with non-vanishing matrix elements can be taken as~\cite{Beneke:2003pa}
\begin{eqnarray}
 [{\rm objects}] \times (\bar \xi_c    \{1,n\!\!\!\slash_+/2\} \Gamma_k' q_s ) (\bar \xi_c     \{1,n\!\!\!\slash_+/2\} \Gamma_l' q_s ) (\bar \xi_c \Gamma_j' h_v).  \label{eq:genericform}
\end{eqnarray}
where the objects in the brackets are combinations of the building blocks:
 \begin{equation}
{\small
\begin{tabular}{|c|c|c|c|c|c|c|c|c|c|cc}   
$(in_-\partial)^{-1} $   & $n_-^\mu  $  & $\partial_\perp, \,A_{\perp c}, \,A_{\perp s} $  &  $n_- \partial,\,n_- A_c $  &  $\bar q_s \frac{ n\!\!\!\slash_+}{2}\Gamma_m^\prime q_s$ & $\bar q_s \Gamma_m^{\prime\prime} q_s$\\   
   $n_1$ & $n_3$ & $n_5$&  $n_7$   & $n_{9a}$ &$ n_{9c}$
  \\\hline
$(in_+\partial)^{-1} $  &      $n_+^\mu  $ &    $n_+ \partial, \,n_+ A_{s} $  &  
$\bar\xi_c \frac{ n\!\!\!\slash_+}{2}\Gamma_m^\prime\xi_c$    &  
$\bar q_s \frac{ n\!\!\!\slash_-}{2}\Gamma_m^\prime q_s$   &  \\ 
 $n_2$ & $n_4$ &$n_6$ & $n_{8}$ & $ n_{9b}$ & \\
\end{tabular} ,
}\nonumber
\end{equation}
with the integers $n_i$ being the number of occurrences of $O_i$ in an operator.  $\Gamma_{j,k,l,m}'$ take one of the forms in Eq.~\eqref{eq:Lorentz-structures}, while $\Gamma_{m}''$ is a basis for the remaining eight boost-invariant Dirac structures.  We,  following Ref.~\cite{Beneke:2003pa}, use the power scaling, boost invariance and the matching of   mass dimensions to pick up the allowed forms.  The notation for these symbols is used as: $[\lambda]_O=n$ means that $O$ scales with $\lambda^n$,  the 
``boost'' label corresponds to  the scaling $\alpha^n$ of $O$ under boosts $n_-\to\alpha n_-$,  
$n_+\to\alpha^{-1}n_+$; the mass dimension is denoted  by $[d]_O$.
Using the  properties of  these  building blocks which are discussed in detail in table 2 of  Ref.~\cite{Beneke:2003pa},  we find  an operator in the final effective theory  has the scalings
\begin{eqnarray}
 &&[\lambda]= 15-2 n_1+2 n_5+2 n_6+ 4 n_7+4 n_8 
+6(n_{9a}+ 
n_{9b}+n_{9c}),  \nonumber\\
 &&[\alpha]=0=  -n_1+n_2+n_3-n_4-n_6+n_{7} 
-n_8-n_{9a}+n_{9b},\nonumber\\
&& [{d }]= 9-n_1-n_2+n_5+n_6+n_{7}
+3(n_8+n_{9a}+n_{9b}+n_{9c}),
\end{eqnarray}
from which we have
\begin{eqnarray}
 [\lambda] = 6+ [d] -n_3+n_4+n_5+2 n_6+2 n_{7}
 +2 n_8+ 4 n_{9a}+ 
2 n_{9b}+3 n_{9c}.
\end{eqnarray}

In operators $O^{A,B,C,E}$ the only Lorentz structure  having nonzero contraction with $n_-^\mu$ is $\epsilon_{\alpha_\perp \beta_\perp\mu\nu} n_-^{\mu} n_+^{\nu}$, and thus $n_3\le n_4$.
For  the $O^A$ operator, $[d]=3$ and there is  only one  nontrivial solution with $[\lambda]=9$, $n_1=n_2=3$, $n_3=n_4$ and $n_i=0(i\ge 5)$.  Since there is no free Lorentz index containing $n_-^\mu$ or $n_+^\mu$ in  $\Gamma_{j}'$, the equality $n_3=n_4$ rules out the possibility of $n\!\!\!\slash_+/2$ in Eq.~\eqref{eq:genericform}.

As  for $O^B$ operator, $[d]=4$ thus $[\lambda]\ge 10$. Since $n_1$ is an integer,  the leading contribution from this operator has the scaling $[\lambda]=11$.  One solution is  $n_1=2$, $n_2=3$, $n_4-n_3=1$,  $n_i=0(i\ge 5)$ and the other is $n_3=n_4$, $n_1=n_2=3$, $n_5=1$, $n_i=0(i\ge 6)$. The latter one corresponds to the higher Fock state contribution, due to the presence of an extra soft or collinear gluon. The $O^E$ also belongs to  this type.

In the light-cone gauge $ (W_{hc}^\dagger in_-D W_{hc} -in_-D_s)$ reduces to $n_-A_{hc}$. 
In the $O^{D}$ operator,  $[d]=4$ and the factor $n_-^\mu$ contracts with the gluon field $A_{hc\mu}$. After the elimination of the hard-collinear fields, $n_-^\mu$ can not be a free Lorentz index and maybe it is contracted: with  $n_+$ which is a constant or  in the form of $\epsilon_{\alpha_\perp \beta_\perp\mu\nu} n_-^{\mu} n_+^{\nu}$; with a gamma matrix as $\bar q_s \frac{ n\!\!\!\slash_-}{2}\Gamma_m^\prime q_s$;  with a derivative to a soft field in the form of $n_-\partial$; or with a collinear gluon field as $n_-A_c$. 
In the first contraction,  $n_4\ge n_3$, and $[\lambda]\ge 10$. Due to the integer constraint on $n_1$,  this operator has the scaling $[\lambda]=11$ and its solution is similar to the one in $O^{B}$. For the rest cases,  $n_3$ may be larger than $n_4$ by one unit, but  $n_7>0$ or $n_{9b}>0$, causing more power suppressions and resulting in $[\lambda]> 11$.

For the  operator $O^C$, $[d]=5$ and  $[\lambda]\ge 11$. The solution having the  power $[\lambda]= 11$  is $n_1=n_2=2$, $n_3=n_4$ and $n_i=0(i \ge 5)$.

The above matching analysis indicates that the operator $O^A$ is indeed dominant and others are $\lambda^2$ suppressed. 
Taking into account the power scalings of baryonic states, we obtain the   scaling laws for   operator matrix elements
\begin{eqnarray}
 \langle \Lambda| O^A|\Lambda_b\rangle \sim \lambda^4,\;\;\;\;
  \langle \Lambda| O^{B,C,D,E}|\Lambda_b\rangle \sim \lambda^6. 
\end{eqnarray}

\subsection{Tree-level Matching}

Now we will perform a  tree-level matching   from SCET$_{I}$ to SCET$_{II}$, and identify various terms to different types of operators.  In this procedure,  the hard-collinear quark field is first  expressed as a product  of soft and collinear fields and the hard-collinear gluon fields. Then the hard-collinear gluons are integrated out by solving the equation of motion for the Yang-Mills fields and their expressions  in terms of    soft and collinear quarks and gluons will be substituted back into the hard-collinear quark field.  
For simplicity, we shall  work in  the light-cone gauge $n_+A_{hc}=n_{+}A_{c}=n_-A_s=0$  and the gauge invariant form can be obtained by the field redefinition technique.

The QCD currents can be matched onto the effective currents in the SCET
\begin{eqnarray}
 J^{QCD} 
= [\bar \psi(x) \Gamma b](x) \to e^{-im_b v\cdot x} [\bar \psi \Gamma {\cal Q}](x)
\end{eqnarray}
with 
\begin{eqnarray}
&&\psi= \xi_c +\eta_c +\xi_{hc} +\eta_{hc}+q_s \nonumber\\
 &&= \xi_c+\xi_{hc}+q_s
 - \frac{1}{ in_+D_s} \frac{n\!\!\!\slash_+}{2} [ (iD\!\!\!\!\slash_{\perp} ) (\xi_c+\xi_{hc}) + (gA\!\!\!\slash_{\perp c} +gA\!\!\!\slash_{\perp hc} )q_s],\nonumber\\
&& {\cal Q}= \left(1+\frac{iD\!\!\!\!\slash_s}{2m_b}\right) h_v -\frac{1}{n_-v} \frac{n\!\!\!\slash_-}{2m_b} (gA\!\!\!\slash_{\perp c} +gA\!\!\!\slash_{\perp hc} )h_v 
\nonumber\\ &&
+ \frac{1}{2m_b n_-v } \left[\frac{1}{in_+\partial} (gA\!\!\!\slash_{\perp c}+ gA\!\!\!\slash_{\perp hc} )(gA\!\!\!\slash_{\perp c} +gA\!\!\!\slash_{\perp hc} ) \right]h_v 
\nonumber\\&& 
  -\left[\left\{\frac{1}{ m_b n_-v}  \frac{n\!\!\!\slash_- n\!\!\!\slash_+}{4}- \frac{ n_+v}{ n_-v i n_+ \partial}  \right\} (n_-A_{hc}) \right]h_v +{\cal O} (\lambda ^4 h_v),\label{eq:treelevel}
\end{eqnarray}
with the derivative $1/n_+\partial$   acting on the collinear field in the square bracket.

 In the light-cone gauge, the collinear quark Lagrangian reads as
\begin{eqnarray}
 {\cal L}= \bar \xi \left( in_-D + [iD\!\!\!\!\slash_\perp ] \frac{1}{in_+D_s}  [iD\!\!\!\!\slash_\perp ]\right) \frac{n\!\!\!\slash_+}{2} \xi +...,
\end{eqnarray}
with the ellipses standing for all other terms. 
Here $\xi$ and the collinear gluon in the covariant derivative  denote  both collinear and hard-collinear field and will be substituted  as $\xi\to \xi_{c}+\xi_{hc}$ $A_{c} \to A_{c}+A_{hc}$. With the use of the equation of motion, the $\xi_{hc}$ can be integrated out and in particular,  the solution (dropping the terms not satisfying momentum conservation) 
\begin{eqnarray}
 \xi_{hc}\sim -\frac{1}{ in_-\partial}   \left( g n_-A_{hc} + iD\!\!\!\!\slash_\perp \frac{1}{ in_+\partial} iD\!\!\!\!\slash_\perp\right)\xi_c \nonumber
\end{eqnarray}
 contributes to $\psi^{(6)}$ as 
\begin{eqnarray} 
 \psi^{(6)}&=& - \frac{1}{ in_-\partial}   \left(  gA\!\!\!\slash^{(3)}_{\perp hc}\frac{1}{ in_+\partial}gA\!\!\!\slash^{(3)}_{\perp hc} + g n_-A_{hc}^{(6)}\right)  \xi_c+... ~, \label{eq:psi6xi6}
\end{eqnarray} 
in which the expressions of gluons will be specified below. 
The other useful pieces are~\cite{Beneke:2003pa}
\begin{eqnarray}
 \psi^{(2)}&=&\xi_c,\nonumber\\
 \psi^{(5)}&=& 
\frac{1}{in_+\partial}\,g A\!\!\!\slash_{\perp hc}^{(3)}\,
  \frac{  n\!\!\!\slash_+}{2}\,\xi_c
-\frac{1}{in_-\partial}\,\bigg((i  D\!\!\!\!\slash_{\perp c}+
  g  A\!\!\!\slash_{\perp s}) 
  \,\frac{1}{in_+\partial}\,g  A\!\!\!\slash_{\perp hc}^{(3)}  \nonumber\\
&&~+ gA\!\!\!\slash^{(3)}_{\perp hc} \frac{1}{ in_+\partial} (iD\!\!\!\!\slash_{\perp c} +gA\!\!\!\slash_{\perp s}) \bigg) \xi_c -\frac{1}{in_-\partial} gn_{-}A_{hc}^{(5)}\xi_c+..., 
\end{eqnarray}
where the first term in $\psi^{(5)}$ is from the small component of the hard-collinear quark field $\eta^{(5)}_{hc}$ and the ones in the large parentheses are from $\xi_{hc}^{(5)}$. 
The relevant hard-collinear gluon field is expanded as~\cite{Beneke:2003pa}
\begin{eqnarray}  
 A_{\perp hc}^{(3)}&=& gT^A \frac{1}{ in_+\partial in_-\partial} \{ \bar q_s \gamma_\perp T^A \xi_c+ h.c. \},\nonumber\\ 
n_- A_{hc}^{(5)} &=& -\frac{2}{(in_+\partial)^2}\,\Bigg\{
  i{\cal D}^{\mu_\perp} [in_+\partial A_{\mu_\perp hc}^{(3)}] - 
  g\,\Big[in_+\partial A^{\mu_\perp}_c,  A_{\mu_\perp hc}^{(3)}\Big]
\nonumber \\ 
&&-\,2g T^A\bigg\{\bar\xi_cT^A\bigg(\frac{  n\!\!\!\slash_+}{2}-
  \frac{1}{in_- \partial}\,g  A\!\!\!\slash_{\perp c}\bigg) q_s+\mbox{h.c.}
  \bigg\}\!\Bigg\},\nonumber\\
 n_-A_{hc}^{(6)}&=& -\frac{2}{ (in_+\partial)^2} \Big[ -2 [in_+\partial A_{\perp hc}^{(3)\mu}, A_{\mu\perp hc}^{(3)} ]  
+2 gT^A  \Big\{\bar \xi_c T^A \Big(\frac{1}{in_-\partial} gA\!\!\!\slash_{\perp hc}^{(3)} \Big) q_s +h.c. \Big\}\Big].
\end{eqnarray} 
with the covariant derivative $i{\cal D}^\mu {\cal  O}= i\partial^\mu {\cal O} + g[A^{\mu}_c+A^{\mu}_s, {\cal O}]$.

Before substituting the hard-collinear fields into the currents, we first count the collinear quark numbers. The final baryonic state  contains three quarks, and has collinear quark number +3.   In order to have nonzero matrix elements, the effective currents in the SCET must   have the collinear quark number 3 as well.  Let us recall that  the gluon filed  $A^{(3)}_{\perp hc}$  contains one collinear quark (or antiquark depending on the interaction form   in the effective theory), while $ n_-A_{hc}^{(6)}$ may contain two collinear quarks.    For the expression of $\psi^{(n)}$,  we note that $\psi^{(2)}$  (and also $\psi^{(4)}$) has collinear quark number $-1$, while $\psi^{(3)}$ has collinear quark number 0. The most nontrivial terms are: $\psi^{(5)}$ which has a collinear quark number $-2$ or 0, and $\psi^{(6)}$ with collinear quark number $-3$ (or $\pm1$).  The combinations having the  leading and next-to-leading power scalings indeed  take the forms as $O^A$, $O^{B,C,D}$ and $O^E$. 

\begin{figure}\begin{center}
\includegraphics[scale=0.5]{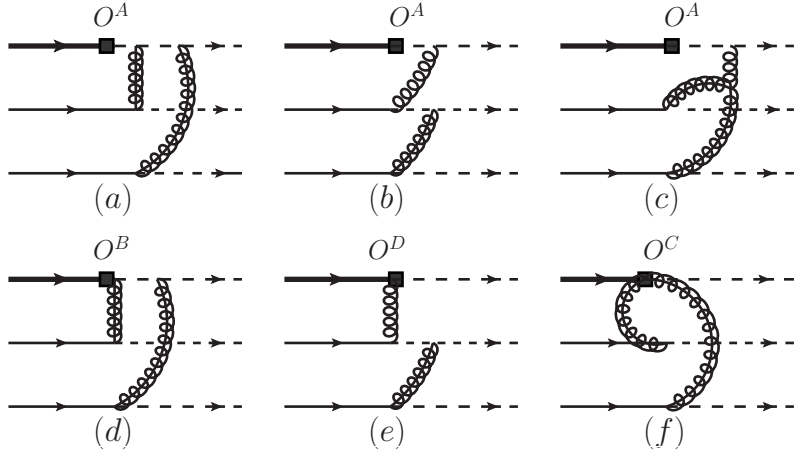}
\caption{Tree matching diagrams for the heavy-to-light baryonic form factors. The dashed lines denote the collinear quarks, while the solid lines are soft spectators. The thick lines represent the heavy bottom quark.  The spring lines denote a collinear gluon $n_-A_{hc}$ while the spring+solid lines denote the $A_{\perp hc}$. The hard modes have been integrated out and shrunk to the black point.    } \label{fig:SCETFF}
\end{center}
\end{figure}

Substituting $A^{(3)}_{\perp hc}$, $n_{-}A_{hc}^{(6)}$ and $\psi^{(6)}$ into the effective currents, we have the leading term in the expansion
\begin{eqnarray}
 &&J^{(9)}=  -\bar \xi_c  \left(  gA\!\!\!\slash^{(3)}_{\perp hc}\frac{1}{- in_+\overleftarrow{\partial}}gA\!\!\!\slash^{(3)}_{\perp hc} + g n_-A_{hc}^{(6)}\right)\frac{1}{ -in_-\overleftarrow{\partial}}    \Gamma h_v. 
\end{eqnarray}
 The first term in the above equation contains two hard-collinear gluons emitted from the hard-collinear quark, and  is depicted as the first diagram in Fig.~\ref{fig:SCETFF}.
In this figure, the dashed lines denote the collinear quarks, while the solid lines are soft spectators. The thick lines represent the heavy bottom quark.  The spring lines denote a collinear gluon $n_-A_{hc}$ while spring+solid lines denote the $A_{\perp hc}$.  
In the $n_-A_{hc}^{(6)}$, the trigluon term, corresponding to Fig~(\ref{fig:SCETFF}c), vanishes and it can be understood as follows.  The three quarks have antisymmetric colors in both initial and final baryons,  and thus the color rearrangement factor in this diagram is zero
\begin{eqnarray}
 \epsilon^{ijk} \epsilon^{i'j'k'} T^A_{ii'} T^B_{jj'} T^C_{kk'}  f^{ABC}=
 \epsilon^{ikj} \epsilon^{i'k'j'} T^A_{ii'} T^B_{kk'} T^C_{jj'}  f^{ABC}=0.
\end{eqnarray} 
The current $J^{(9)}$ originates from the large component of the hard-collinear quark field $ \xi_{hc}^{(6)}$ as shown in Eq.~\eqref{eq:psi6xi6} and thereby the Lorentz structure is reduced: 
\begin{eqnarray}
J^{(9)}\sim \frac{n\!\!\!\slash_+n\!\!\!\slash_-}{4}\Gamma \frac{1+v\!\!\!\slash}{2}\to \Gamma_{j}', \label{eq:LEETreduction}
\end{eqnarray}
 as expected in the large recoil limit.

The other combinations of operators start  from $\lambda^{11}$
\begin{eqnarray}
 J^{(11)}&=&- \frac{1}{n_-v} \bar \psi^{(5)}  \Gamma \frac{n\!\!\!\slash_-} {2m_b} gA\!\!\!\slash^{(3)}_{\perp hc}h_v  +  \frac{1}{2m_bn_-v} \bar \psi^{(2)}   \Gamma \frac{1}{ in_+\partial}  gA\!\!\!\slash^{(3)}_{\perp hc} gA\!\!\!\slash^{(3)}_{\perp hc} h_v \nonumber\\
 && - 
  \bar \psi^{(2)}   \Gamma \left[\left\{\frac{1}{ m_b n_-v}  \frac{n\!\!\!\slash_- n\!\!\!\slash_+}{4}- \frac{ n_+v}{ n_-v i n_+ \partial}  \right\} (n_-A_{hc}^{(6)}) \right]h_v   + \bar \psi^{(6)}\Gamma \frac{iD\!\!\!\!\slash_s}{2m_b} h_v+...,
\end{eqnarray} 
where these four pieces can be incorporated into the operators $O^{B,C,D,E}$ respectively. It should be noted that except the second term, the other terms can have  different Lorentz structures  with the reduced  form as in Eq.~\eqref{eq:LEETreduction}.  For instance, the fourth term is from the small component of the heavy bottom quark, which has the Lorentz structure $\frac{n\!\!\!\slash_+n\!\!\!\slash_-}{4} \Gamma \frac{1-v\!\!\!\slash}{2} $.

We also show the tree-level  matching diagrams for the $O^{B,C,D}$ operators  in  Fig.~\ref{fig:SCETFF}. However the higher Fock state contributions,  either from $O^E$ having the similar structure with $O^A$ except that one additional soft gluon is emitted from the hard vertex,  or from the operator $O^B$,  are not   depicted.   Graphically speaking the dominance of $O^A$ can be understood as follows. In the three diagrams (a,d,e) one commonality is that the two gluons interact with a soft quark from the initial state and a collinear quark in the final external state, and thereby these two vertices have the same power scaling.  However in the first diagram the quark propagator next to the weak vertex  has the form $1/(n_-p)\sim 1/\lambda^2$ while the rest quark propagators are of order ${\lambda^0}$, leading to the enhancement of the first diagram. 


\section{ Analysis of the transition diagrams in QCD}
\label{eq:QCDanalysis}

\begin{figure}\begin{center}
\includegraphics[scale=0.5]{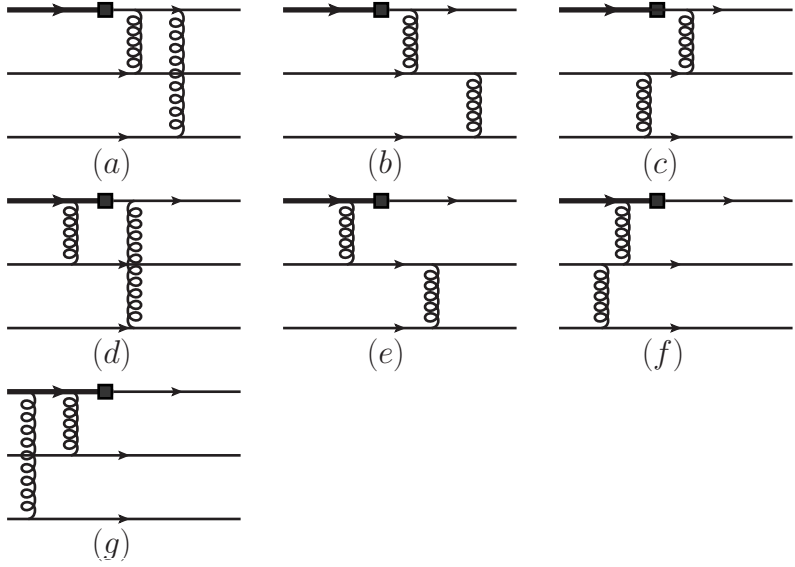}
\caption{Feynman Diagrams for heavy-to-light baryonic form factors in QCD. Trigluon diagrams having wrong color factors are not shown.  } \label{fig:QCDFF}
\end{center}
\end{figure}

In this section, 
we will analyze the leading power behaviors of  the  baryonic transition form factors in QCD, whose Feynman diagrams are depicted in  Fig.~\ref{fig:QCDFF}. We adopt the ansatz that the   fast-moving baryon is composed of three collinear constituents, therefore at least two gluons are exchanged and  these gluons must be far off-shell. We will  not include the contributions involving higher Fock states,  as at least one more gluon is needed.   As we  have already shown, 
the trigluon diagrams  give vanishing contributions and  thereby will not be considered either. 
There are seven diagrams shown in  Fig.~\ref{fig:QCDFF}:  three of them (a,b,c)  containing the momentum exchange by two gluons between the spectator quark system  and the energetic light quark connecting the the electroweak vertex;  the same number of diagrams (e,f,g) having two gluons emitted from the heavy quark;  the rest diagram (d) in which the light spectator system receives momentum exchange from both  the energetic quark and the heavy quark.  The inclusion of the flavor index will give another seven diagrams, but only leads to the exchange of momentum fractions of the light spectator quarks.

The leading twist LCDA  of a light baryon, such as $\Lambda$,  is~\cite{arXiv:0811.1812,Mannel:2011xg}
\begin{eqnarray}
  \epsilon^{ijk} \frac{ n_+p'}{8} \frac{1}{6} (Cn\!\!\!\slash_- \gamma_5)_{\beta\alpha} (\bar u_{\Lambda})_{\gamma}, 
\end{eqnarray} 
with $i,j,k$ being the color indices and $\alpha,\beta,\gamma$ being the spinor indices. 
For the heavy baryon, several types of LCDAs emerge~\cite{arXiv:0804.2424}
\begin{eqnarray} 
\frac{1}{48}  \epsilon^{ijk} (n\!\!\!\slash_+ \gamma_5C)_{\alpha\beta} (u_{\Lambda_b})_{\gamma},  \frac{1}{48}
  \epsilon^{ijk} (n\!\!\!\slash_- \gamma_5C)_{\alpha\beta} (u_{\Lambda_b})_{\gamma},  \nonumber\\ \frac{1}{48}\epsilon^{ijk} ( \gamma_5C)_{\alpha\beta} (u_{\Lambda_b})_{\gamma},   \frac{1}{48}
  \epsilon^{ijk} (n\!\!\!\slash_+ n\!\!\!\slash_- \gamma_5C)_{\alpha\beta} (u_{\Lambda_b})_{\gamma}. 
\end{eqnarray}
In the leading power matrix elements, only the first type of LCDA contributes. 
We choose the momentum fractions of the three collinear quarks in the light baryon as $y_1,y_2,y_3$  and the momentum fractions of the soft spectator quarks (in the direction $n_+$)  in the initial state as $x_2$ and $x_3$. The corresponding momenta will be denoted as $p_1',p_2',p_3'$ for the   collinear quarks, and $p_2,p_3$ for the soft quarks.

The calculation will be simplified by the following two observations. 
\begin{itemize}
\item  If both  vertices of a hard-collinear gluon are attached to collinear quarks, only the transverse component of this  gluon contributes.  

\item In the light spectator system (usually called a diquark), only the diagrams with even number of gluon transverse indices   are nonzero. For instance,  as shown in Fig~\ref{fig:QCDFF}(b,e,f), if only one hard-collinear gluon is emitted from the light quark or the heavy quark,   this gluon has to be in the form  $n_-A_{hc}$ or $n_+A_{hc}$.  
\end{itemize}
The first observation can be proved by writing    the amplitudes   as
\begin{eqnarray}
 [\bar q_1 \gamma_{\mu} ...] \times 
 [\bar q_2 \gamma^{\mu} ...] = 
 [\bar q_1 \gamma_{\perp\mu} ...] \times 
 [\bar q_2 \gamma^{\perp\mu} ...],
\end{eqnarray}
with $q_1$ and $q_2$ being the two collinear quarks attached to the gluon. 
The second one is based on the fact that   the two-spectator system  technically forms  a trace in the spinor space. There is no transverse index from the external wave functions, and thereby the internal ones from the exchanged gluons must be even. 

The leading power contributions from 
Fig.~\ref{fig:QCDFF}(a,b) can be matched onto the $O^A$ operator. 
In Fig.~\ref{fig:QCDFF}(a),  using  the first observation, one of the two gluons (the right one) is connected to two collinear quarks, and only the transverse component is left.  With the second observation, the other gluon must    take the transverse component as well. In the numerator of  the quark propagator between the two gluons,  the collinear momentum $ p\!\!\!\slash_1'+p\!\!\!\slash_3'$ does not contribute since it is next to the light spinor: $\bar u_{\Lambda} \gamma_\perp (p\!\!\!\slash_1'+p\!\!\!\slash_3')=0$. This propagator is simplified as
\begin{eqnarray}
 \frac{i (-p\!\!\!\slash_3+ p\!\!\!\slash_1'+p\!\!\!\slash_3')}{ (-p_3+p_1'+p_3')^2}\simeq \frac{ in\!\!\!\slash_+ }{2 (y_1+y_3) n_+p'},
\end{eqnarray}
which scales as $\lambda^0$. The other quark propagator is reduced to
\begin{eqnarray}
 \frac{i (-p\!\!\!\slash_3-p\!\!\!\slash_2+ p\!\!\!\slash')}{ (-p_2-p_3+p')^2}\simeq -\frac{ in\!\!\!\slash_- }{2 (x_2+x_3) m_{\Lambda_b} n_-v},\label{eq:2propagator}
\end{eqnarray}
which  has the scaling   $1/\lambda^2$. Here we have used $x_2, x_3\sim \Lambda/m_{\Lambda_b}$ for the soft momentum fraction. Combining these pieces,  this  diagram has the form 
\begin{eqnarray}
 F^{(a)}&= &C_N g_s^4 \int dy_2dy_3 dx_2 dx_3 f_{\Lambda_b} f_{\Lambda} \Phi_{\Lambda_b} (x_1,x_2,x_3)\Phi_{\Lambda} (y_1,y_2,y_3)  
\nonumber\\&&  \times 
 \frac{i}{y_3 x_3 m_{\Lambda_b} n_+p'n_-v }\frac{i}{y_2 x_2 m_{\Lambda_b} n_+p' n_-v} \nonumber\\
&& \times \bar u_{\Lambda} \gamma_{\perp }^\mu \frac{ in\!\!\!\slash_+ }{2 (y_1+y_3) n_+p' }\gamma_{\perp }^\nu \frac{- in\!\!\!\slash_- }{2 (x_2+x_3) m_{\Lambda_b} n_-v}\Gamma u_{\Lambda_b} \nonumber\\
&& \times  \frac{ n_+p'}{64}  (C n\!\!\!\slash_- \gamma_5)_{\alpha\beta}  (\gamma_{\nu})_{\alpha\alpha'}  (\gamma_{\mu})_{\beta\beta'} (n\!\!\!\slash_+ \gamma_5C)_{\beta' \alpha'}  
\nonumber\\
&  \propto& \lambda^{10}/\lambda^6 \sim \lambda^4,\label{eq:figureaQCD}
\end{eqnarray}
where the scaling $\lambda^{10}$ is from  decay constants  and $1/\lambda^6$ comes from the two gluons propagators and the propagator in Eq.~(\ref{eq:2propagator}).   $C_N$ is the color factor 
\begin{eqnarray}
 C_N = \frac{1}{36} \epsilon^{ijk} \epsilon^{i'j'k'} (T^bT^a)^{ii'} (T^b)^{jj'} (T^a)^{kk'}=  \frac{2}{27}. 
\end{eqnarray} 
Eq.~\eqref{eq:figureaQCD}   confirms our power counting analysis given in the previous section.  Furthermore as indicated in the third line of the above equation,   the light spectator (diquark) system is proportional to $g^{\perp \mu\nu}$ which results in the Lorentz structure  
\begin{eqnarray}
 F^{(a)}&\propto &   \bar u_{\Lambda} \frac{ n\!\!\!\slash_+ n\!\!\!\slash_-}{4}\Gamma \frac{1+v\!\!\!\slash}{2} u_{\Lambda_b},
\end{eqnarray}
where the large energy and heavy quark symmetries are manifestly demonstrated again.

In Fig.~\ref{fig:QCDFF}(b),  the  upper gluon vertex is replaced by $  n\!\!\!\slash_+/2$ and the quark propagator next to the electroweak vertex is of the form $  n\!\!\!\slash_-/2$. Therefore this diagram has the same structure:
\begin{eqnarray}
 F^{(b)}&= &C_N g_s^4  \int dy_2dy_3 dx_2 dx_3 f_{\Lambda_b} f_{\Lambda} \Phi_{\Lambda_b} (x_1,x_2,x_3)\Phi_{\Lambda} (y_1,y_2,y_3)  
\nonumber\\
&& \times 
 \frac{i}{y_3 x_3 m_{\Lambda_b} n_+p' n_-v }\frac{i}{(y_2+y_3)( x_2+x_3) m_{\Lambda_b} n_+p' n_-v }
\times \bar u_{\Lambda}    \frac{  n\!\!\!\slash_+ }{2}   \frac{- in\!\!\!\slash_- }{2(x_2+x_3)  n_-v m_{\Lambda_b}}\Gamma u_{\Lambda_b}   \nonumber\\
&& \times \frac{ n_+p'}{2}  (C n\!\!\!\slash_- \gamma_5)_{\alpha\beta}  (\gamma_{\perp \mu} \frac{ in\!\!\!\slash_+ }{2 (y_2+y_3) n_+p'}   n\!\!\!\slash_-  )_{\alpha\alpha'}  (\gamma_\perp^{\mu})_{\beta\beta'} (n\!\!\!\slash_+ \gamma_5C)_{\beta'\alpha'}
\nonumber\\
&  \propto& \lambda^{10}/\lambda^6 \sim \lambda^4. 
\end{eqnarray}  
In Fig.~\ref{fig:QCDFF}(c), based on the first observation,  the upper gluon is transverse;  thus there are either one or three transverse indices in the light spectator system, leading to vanishing contribution. 

In 
Fig.~\ref{fig:QCDFF}(d),  both gluons can only contain transverse components and this diagram can be matched onto  the operator $O^{B}$.  Both the heavy quark and light quark propagators   scale as $\lambda^0$ and thus 
\begin{eqnarray} 
 F^{(d)}&= &C_N g_s^4 \int dy_2dy_3 dx_2 dx_3 f_{\Lambda_b} f_{\Lambda} \Phi_{\Lambda_b} (x_1,x_2,x_3)\Phi_{\Lambda} (y_1,y_2,y_3)  
 \frac{i}{y_3 x_3 m_{\Lambda_b} n_+p' n_-v}\frac{i}{x_2y_2 m_{\Lambda_b} n_+p' n_-v} \nonumber\\
&& \times \bar u_{\Lambda} \gamma_{\perp}^\nu     \frac{ in\!\!\!\slash_+}{2(y_1+y_3)  n_+p'}\Gamma  \frac{ in\!\!\!\slash_-}{2m_{\Lambda_b} n_-v } \gamma_{\perp}^\mu u_{\Lambda_b}  \times \frac{ n_+p'}{2}  (C n\!\!\!\slash_- \gamma_5)_{\alpha\beta}  (\gamma_{\perp \mu}  )_{\alpha\alpha'}  (\gamma_{\perp\nu})_{\beta\beta'} (n\!\!\!\slash_+ \gamma_5C)_{\beta'\alpha'}
\nonumber\\
&  \propto& \lambda^{10}/\lambda^4 \sim \lambda^6. 
\end{eqnarray}   
with again $\lambda^{10}$ from decay constants and $1/\lambda^4$ from the two gluon propagators. Of particular interest is that the Lorentz structure in this diagram has the form
\begin{eqnarray}
 F^{(d)}&\propto & \bar u_{\Lambda} \gamma_{\perp}^\nu     \frac{ in\!\!\!\slash_+}{2(y_1+y_3)  n_+p'}\Gamma  \frac{ in\!\!\!\slash_-}{2m_{\Lambda_b} n_-v }\gamma_{\perp}^\mu u_{\Lambda_b},
\end{eqnarray}
which manifestly breaks the large recoil symmetries. 

In Fig.~\ref{fig:QCDFF}(e), the gluon attaching to the two light quarks is transverse while the component $n_{-} A_{hc}$ contributes at the heavy quark propagator. This diagram corresponds  to the operator $O^{D}$. Using the first observation, the light quark propagator scales as $\lambda^0$ and thus
\begin{eqnarray} 
 F^{(e)}&= &C_N g_s^4 \int dy_2dy_3 dx_2 dx_3 f_{\Lambda_b} f_{\Lambda} \Phi_{\Lambda_b} (x_1,x_2,x_3)\Phi_{\Lambda} (y_1,y_2,y_3)  
\nonumber\\
&& \times 
 \frac{i}{y_3 x_3 m_{\Lambda_b} n_+p' n_-v }\frac{i}{(y_2+y_3)( x_2+x_3) m_{\Lambda_b} n_+p' n_-v }
  \nonumber\\
&& \times \bar u_{\Lambda}  \Gamma  \left[ \frac{- i n_+v } {(y_2+y_3) n_+p' n_-v}+  \frac{ n\!\!\!\slash_- n\!\!\!\slash_+ }{4}  \frac{ i }{ n_-v m_{\Lambda_b}}  \right] u_{\Lambda_b}   \nonumber\\
&& \times \frac{ n_+p'}{2}  (C n\!\!\!\slash_- \gamma_5)_{\alpha\beta}  (\gamma_{\perp \mu} \frac{ in\!\!\!\slash_+ }{2 (y_2+y_3) n_+p'}   n\!\!\!\slash_-   )_{\alpha\alpha'}  (\gamma_\perp^{\mu})_{\beta\beta'} (n\!\!\!\slash_+ \gamma_5C)_{\beta'\alpha'}
\nonumber\\
&  \propto& \lambda^{10}/\lambda^4 \sim \lambda^6.   
\end{eqnarray}
The first term in the square bracket obeys the large recoil symmetries, but the integral over $y_2+y_3$ in it is divergent. 
It is worthwhile to point out that in the SCET solution for the operator $O^E$ in the previous section, the number of the occurrence of $1/(in_+\partial)$ is found to be $n_2=3$, which means the momentum fractions for the light baryon can appear only three times. The additional momentum fraction   arises from the short-distance coefficients, for instance at tree-level shown in Eq.~\eqref{eq:treelevel}. 

In Fig.~(\ref{fig:QCDFF}f),  the gluon attaching to the $b$ quark can not contribute with the transverse component based on the second observation. The $n_+A_{hc}$ component can be absorbed into the Wilson line, one necessary piece in the gauge invariant definition of the SCET operators. Thus this diagram is incorporated into the operator $O^A$ and its scaling is
\begin{eqnarray} 
 F^{(f)}&= &C_Ng_s^4 \int dy_2dy_3 dx_2 dx_3 f_{\Lambda_b} f_{\Lambda} \Phi_{\Lambda_b} (x_1,x_2,x_3)\Phi_{\Lambda} (y_1,y_2,y_3)  
\nonumber\\
&& \times 
 \frac{i}{y_3 x_3 m_{\Lambda_b} n_+p' n_-v }\frac{i}{(y_2+y_3)( x_2+x_3) m_{\Lambda_b} n_+p' n_-v }
\times \bar u_{\Lambda}  \Gamma   \frac{- i} {(y_2+y_3) n_+p' } u_{\Lambda_b}  
\nonumber\\
&& \times  \frac{ n_+p'}{64}  (Cn\!\!\!\slash_- \gamma_5 )_{\alpha\beta}  (  n\!\!\!\slash_+  \frac{- in\!\!\!\slash_- }{2 (x_2+x_3) m_{\Lambda_b} n_-v } \gamma_{\perp\mu}   )_{\alpha\alpha'}  (\gamma_\perp^{\mu})_{\beta\beta'} (n\!\!\!\slash_+ \gamma_5C)_{\beta'\alpha'}
\nonumber\\
&  \propto& \lambda^{10}/\lambda^6 \sim \lambda^4.   
\end{eqnarray}
In particular this contribution cancels the one from Fig.~(\ref{fig:QCDFF}b).

In Fig.~\ref{fig:QCDFF}(g), the two heavy quark propagators have the offshellness of order $m_b^2$ and  can be shrunk to one point. Suppose that the two gluons are transverse,  and then it is incorporated into   $O^C$ and its power scaling is 
\begin{eqnarray} 
 F^{(g1)}&= &C_N g_s^4 \int dy_2dy_3 dx_2 dx_3 f_{\Lambda_b} f_{\Lambda} \Phi_{\Lambda_b} (x_1,x_2,x_3)\Phi_{\Lambda} (y_1,y_2,y_3)   
 \frac{i}{y_3 x_3 m_{\Lambda_b} n_+p' n_-v }\frac{i}{x_2 y_2m_{\Lambda_b} n_+p' n_-v }
  \nonumber\\
&& 
\times \bar u_{\Lambda}  \Gamma   \frac{ -1} {(y_2+y_3) n_+p' }\gamma_{\perp \mu} \gamma_{\perp \nu} u_{\Lambda_b}  
\  \times  \frac{ n_+p'}{2}  (C n\!\!\!\slash_-\gamma_5)_{\alpha\beta}  ( \gamma_{\perp\mu}   )_{\alpha\alpha'}  (\gamma_\perp^{\mu})_{\beta\beta'} (n\!\!\!\slash_+ \gamma_5C)_{\beta'\alpha'}
\nonumber\\
& \propto& \lambda^{10}/\lambda^4 \sim \lambda^6,
\end{eqnarray}
where the momentum fraction $1/(y_2+y_3)$ in the second line comes   from the Wilson coefficient for the operator $O^C$. 
If the two gluons take the $n_{+ }A_{hc}$ component for the vertices attaching to the heavy quark, this diagram can be matched   onto operator $O^A$. The contribution is $\lambda^2$ suppressed compared to the leading power terms 
\begin{eqnarray} 
 F^{(g2)}&= &C_Ng_s^4 \int dy_2dy_3 dx_2 dx_3 f_{\Lambda_b} f_{\Lambda} \Phi_{\Lambda_b} (x_1,x_2,x_3)\Phi_{\Lambda} (y_1,y_2,y_3)  
 \times 
 \frac{i}{y_3 x_3 m_b n_+p' n_-v }\frac{i}{x_2 y_2m_b n_+p' n_-v }
  \nonumber\\
&& 
\times \bar u_{\Lambda}  \Gamma   \frac{ i} {(y_2+y_3) n_+p' } \frac{ i} {y_3 n_+p' } u_{\Lambda_b}  
\  \times  \frac{ n_+p'}{2}  (C n\!\!\!\slash_-\gamma_5)_{\alpha\beta}  (  n\!\!\!\slash_+  )_{\alpha\alpha'}  (n\!\!\!\slash_+  )_{\beta\beta'} (n\!\!\!\slash_- \gamma_5C)_{\alpha'\beta'}
\nonumber\\
&  \propto& \lambda^{10}/\lambda^4 \sim \lambda^6, 
\end{eqnarray}
and the integration in this term does not converge.

\section{Discussions}\label{sec:discussions}

As we have shown, in  the dominant contribution from the $O^A$  the inverse of derivatives to both collinear fields and soft fields appear    three times. In the momentum space these factors will be converted to the inverse of momenta. Let them act  on the collinear fields, we obtain the factor $1/(n_+p')^3$. The energy dependence of a quark field can be read from the propagators
\begin{eqnarray}
 \langle 0|\xi_{c}(x)\bar \xi_c(0)|0\rangle = \int  \frac{d^4p'}{(2\pi)^4}e^{-i n_+p' n_-x/2} \frac{n_+p'  } {p^{\prime 2}}\frac{n\!\!\!\slash_-}{2}. 
\end{eqnarray}  
The measure $d^4p'$ and  $p^{\prime 2}$ are Lorentz invariant, and thus 
$\xi_c\sim \sqrt {n_+p'}$.  Then the leading power baryonic transition matrix element   scales as 
\begin{eqnarray}
 \langle\Lambda (p')|O^A(0) |\Lambda_b(p)\rangle \sim \frac{{(n_+p')}^{3/2}} {{n_+p'}^{3}} =   {(n_+p')}^{-3/2}, \nonumber
\end{eqnarray} 
where we have employed the energy independence of baryon states.  
Using $\bar u_{\Lambda}\sim \sqrt{ n_+p'}$ and the definition of the soft form factor in SCET 
\begin{eqnarray}
 \langle\Lambda (p')|O^A(0) |\Lambda_b(p)\rangle =   \xi_\Lambda (E)\bar u_{\Lambda}(p')\Gamma u_{\Lambda_b}(p),\nonumber
\end{eqnarray} 
and restoring correct mass dimensions, we obtain the momentum dependence  
\begin{eqnarray}
  \xi_{\Lambda} (E)\sim \frac{\Lambda^2}{(n_+p')^2}. 
\end{eqnarray}
This behavior can also be read from the QCD calculation as shown in Eq.~\eqref{eq:figureaQCD}. But it should be noticed that  the above scaling law is different with the results derived in different versions of  QCD  light-cone sum rules~\cite{Feldmann:2011xf,Mannel:2011xg}  in which the form factor is dominated by soft processes. To have the power counting, we represent the form factor as an overlap integral of the wave functions in both longitudinal and transverse momentum space
\begin{eqnarray}
 \xi_{\Lambda} (E)= \int \frac{dx_2 d^2 \vec k_{2\perp} dx_3 d^2 \vec k_{3\perp} } {(16\pi^3)^2} 
\psi_{\Lambda_b}( x_2,x_3,\vec k_{2\perp}, \vec k_{3\perp}) \psi_{\Lambda}( y_2(x_2), y_3(x_3),\vec k_{2\perp}, \vec k_{3\perp}),
\end{eqnarray}
with $y_{2}(x_2)$ and $y_3(x_3)$ to be fixed by kinematics. From the normalizations of the b-baryon state, we have 
\begin{eqnarray} 
\int \frac{dx_2 d^2 \vec k_{2\perp} dx_3 d^2 \vec k_{3\perp} } {(16\pi^3)^2} 
 |\psi_{\Lambda_b}( x_2,x_3,\vec k_{2\perp}, \vec k_{3\perp})|^2=1,
\end{eqnarray}
implying that $\psi_{\Lambda_b}( x_2,x_3,\vec k_{2\perp}, \vec k_{3\perp})\sim \lambda^{-6}$ since $x_{2,3}\sim \lambda^2$ and $k_{2\perp, 3\perp}\sim \lambda^2$. For the light particles, the momentum fraction in the normalization is of order 1, therefore for generic values of $y_{2,3}$, $\psi_{\Lambda}( y_2, y_3,\vec k_{2\perp}, \vec k_{3\perp})\sim \lambda^{-4}$. However the dominance of soft processes leads to the  phase suppression  and in particular  the scalings of the momentum fractions  $y_2(x_2)\sim \lambda^2$ and $y_3(x_3)\sim \lambda^2$ result in $\psi_{\Lambda}( y_2(x_2), y_3(x_3),\vec k_{2\perp}, \vec k_{3\perp})\sim 1$. Substituting the scalings for the wave-functions, we obtain 
\begin{eqnarray}
 \xi_{\Lambda} (E) \sim \lambda^{6},
\end{eqnarray}
from which we can see  the contribution from the soft process is formally $\lambda^2$-suppressed compared to the leading power contribution from the operator $O^A$.

As a  comparison, it is also instructive to recapture the energy dependence of the $B\to \pi$ form factor  in the SCET.  Ref.~\cite{Beneke:2003pa} finds that when matching onto SCET$_{II}$ the leading power contribution, from the operator 
$\bar \xi_{hc} \Gamma h_v$,  has two powers of $1/(in_+\partial)$. Together with the scalings from the two collinear quark fields,   the soft form factor, parametrized  via 
\begin{eqnarray}
 \langle \pi(p') |\bar \xi_{hc}  h_v |\bar B(p)\rangle =2E \xi_{\pi}(E) \nonumber
\end{eqnarray}
with $E= n_-v n_+p'/2= (m_B^2-q^2)/(2m_B)$, behaves as
\begin{eqnarray}
 \xi_{\pi} (E)\sim \frac{\Lambda^{3/2} \sqrt {m_b}}{ (n_+p')^2} \sim \frac{(\Lambda/m_B)^{3/2}}{ 1-q^2/m_B^2}. 
\end{eqnarray}

When matching to SCET$_{II}$,  the derivatives $1/(in_+\partial)$ and $1/(in_-\partial)$ also contain the momentum fractions: $x_2, x_3$ or $x_2+x_3$ for the initial heavy baryon, $y_1,y_3,y_3$ or some linear combinations depending on the fields acting on. For example, 
the tree-level  factorization formula  from Fig.~\ref{fig:QCDFF}(a) has  the following integration form as shown in Eq.~\eqref{eq:figureaQCD}
\begin{eqnarray}
  \int dy_2dy_3 dx_2 dx_3   \Phi_{\Lambda_b} (x_1,x_2,x_3)\Phi_{\Lambda} (y_1,y_2,y_3)  
  \frac{1}{ x_2x_3y_2 y_3 (y_1+y_3)(x_2+x_3)  } .
\end{eqnarray}
With the assumption that $\Phi \sim x_2 x_3$ in the limit of $x_2,x_3\to 0$~\cite{arXiv:0804.2424,arXiv:0811.1812}, where  $\Phi$ denotes the LCDA  of   $\Lambda_b$ or $\Lambda$,
the integration is convergent which is different with  the mesonic  transition form factor $\xi_\pi$.  In Fig.~\ref{fig:QCDFF}(b) and Fig.~\ref{fig:QCDFF}(f), the involved  integral 
\begin{eqnarray} 
 \int dy_2dy_3 dx_2 dx_3  \Phi_{\Lambda_b} (x_1,x_2,x_3)\Phi_{\Lambda} (y_1,y_2,y_3)   
 \frac{1}{y_3 x_3  (y_2+y_3)^2( x_2+x_3)^2   } 
\end{eqnarray} 
is finite as well.  The absence of the divergences   leads to the factorization of $\xi_{\Lambda}$
\begin{eqnarray}
\xi_{\Lambda}= f_{\Lambda_b} \Phi_{\Lambda_b} (x_i)\otimes  J(x_i,y_i)\otimes f_{\Lambda} \Phi_{\Lambda}(y_i),\label{eq:formfactorfactorization}
\end{eqnarray}
in which $\otimes$ denotes the convolution over momentum fractions $x_i$ and $y_i$, and  the jet function is given as 
\begin{eqnarray} 
 J(x_i,y_i)=  -\frac{1}{4} C_Ng_s^4 \frac{1}{ x_2 x_3(x_2+x_3)} \frac{1}{ y_2 y_3(y_1+y_3)}  \frac{1}{( m_{\Lambda_b}^2-q^2)^2m_{\Lambda_b} }+ (x_2 \leftrightarrow x_3, y_2\leftrightarrow y_3). 
\end{eqnarray}
It should be cautious that although this formula is valid at tree-level (order $\alpha_s^2$), whether it  can be extended to all orders remains unknown to us and  requires further analysis. 

On the contrary, the  subleading power corrections  can not be factorized, for instance the second term from the diagram shown in Fig.~\ref{fig:QCDFF}(g),  has the  form 
\begin{eqnarray}
 \int dy_2 dy_3 \frac{1}{y_2 y_3^2 (y_2+y_3)} \Phi_{\Lambda} (y_1,y_2,y_3) \sim \log (y_3), \nonumber
\end{eqnarray}
which is divergent when $y_3$ is approaching zero.

To have  some numerical  estimate, 
we use the QCD sum rule calculation of the   $f_{\Lambda_b}$(next-to-leading order in $\alpha_s$)~\cite{Groote:1997yr} and  $f_{\Lambda}$~\cite{arXiv:0811.1812}
\begin{eqnarray}
 f_{\Lambda_b}= (0.032\pm 0.004) {\rm GeV}^3,\;\; f_{\Lambda}= (6.0\pm 0.3) \times 10^{-3}  {\rm GeV}^2,
\end{eqnarray}
together with the asymptotic form  of $\Phi_{\Lambda}$~\cite{arXiv:0811.1812} and the parametrized model for $\Phi_{\Lambda_b}$~\cite{arXiv:0804.2424}
\begin{eqnarray}
 \Phi_{\Lambda_b}(x_1,x_2,x_3) = x_2 x_3 \left[ \frac{m_{\Lambda_b}^4}{\epsilon_0^4} e^{-(x_2+x_3)m_{\Lambda_b}/\epsilon_0} +a_2 C_2^{3/2} (2u-1) \frac{m_{\Lambda_b}^4}{\epsilon_1^4} e^{-(x_2+x_3)m_{\Lambda_b}/\epsilon_1}\right],\nonumber\\
 \Phi_{\Lambda}(y_1,y_2,y_3)= 120 y_1y_2y_3,
\end{eqnarray}
where $\omega= (x_2+x_3)m_{\Lambda_b}$, $u= x_2/(x_2+x_3)$, $\epsilon_0= (200^{+130}_{-60})$ MeV, $\epsilon_1= (650^{+650}_{-300})$ MeV and $a_2= 0.333^{+0.250}_{-0.333}$~\cite{arXiv:0804.2424}. With these inputs and the strong coupling constant at the scale $\mu \sim 2$ GeV: $\alpha_s \simeq 0.3$,  we calculate the form factor as 
\begin{eqnarray}
 \xi_{\Lambda}({q^2=0})= -0.012^{+0.009}_{-0.023},
\end{eqnarray}
where the displayed uncertainties are from $\epsilon_0$. For comparison, we quote the soft form factor $\xi_{\Lambda}$ computed  in the SCET sum rules~\cite{Feldmann:2011xf}
\begin{eqnarray}
 \xi_{\Lambda}({q^2=0})=0.38,
\end{eqnarray}
which is  larger by about one order of magnitude.


\section{Conclusions}
\label{sec:conclusion}

Weak decays of heavy baryons provide an ideal ground for the extraction of the helicity structure of the electroweak interaction, thanks to the spin correlation and polarization embedded in   decay amplitudes. In the heavy-to-light transition, the most important ingredients incorporating the QCD dynamics are form factors. Due to the variety in the Lorentz structures, the amplitude is governed by a number of form factors. 
The development of the effective field theory allows us to simplify the form factors and pick up the terms of great importance. 

In this work we have analyzed the factorization properties and power scalings of heavy-to-light baryonic form factors at large recoil. Using the  soft-collinear effective theory, we proved that the form factors are greatly simplified  by the heavy quark and large energy symmetries at leading power in $1/m_b$. This finding indicates that  only one function is necessary to parametrize the transition of $\Lambda_b\to p$ or $\Lambda_b\to \Lambda$.  A general power counting analysis indicates the  form factors are of the order  $\Lambda^2/E^2$. In contrast to  the mesonic case,   the leading power form factor can factorize into a convolution  of a hard-scattering kernel of order $\alpha_s^2$ and  light-cone distribution amplitudes without encountering any divergence.  Using the inputs mainly from QCD sum rules, we calculate the  form factor $\xi_{\Lambda}(E)$  and find it is numerically smaller than the one   governed by soft processes, although the latter is formally power-suppressed. 
We have also discussed the origins for  symmetry breaking effects which are  suppressed by powers of  ${\Lambda}/m_{b}$ and/or  ${\Lambda}/E$.

\section*{ Acknowledgement} 
The author is very grateful to Ahmed Ali for valuable discussions and carefully reading this manuscript,  and to  Thorsten Feldmann for pointing out an error in the earlier  version  of this manuscript.  He also thanks Yu-Ming Wang and De-Shan Yang for useful discussions.  
This work is supported by the Alexander von Humboldt foundation.


\begin{thebibliography}{99} 

\bibitem{arXiv:0801.1833} 
  M.~Artuso, D.~M.~Asner, P.~Ball, E.~Baracchini, G.~Bell, M.~Beneke, J.~Berryhill and A.~Bevan {\it et al.},
  Eur.\ Phys.\ J.\ C\ {\bf 57}, 309  (2008)
  [arXiv:0801.1833 [hep-ph]].

\bibitem{UTPT-90-03} 
  N.~Isgur and M.~B.~Wise,
  Nucl.\ Phys.\ B\ {\bf 348}, 276  (1991).
\bibitem{Mannel:1990vg}
  T.~Mannel, W.~Roberts, Z.~Ryzak,
  Nucl.\ Phys.\  {\bf B355}, 38-53 (1991).

\bibitem{hep-ph/9701399} 
  T.~Mannel and S.~Recksiegel,
  J.\ Phys.\ GG\ {\bf 24}, 979  (1998)
  [hep-ph/9701399].


\bibitem{Feldmann:2011xf}
  T.~Feldmann, M.~W.~Y.~Yip,
  [arXiv:1111.1844 [hep-ph]].


\bibitem{Mannel:2011xg}
  T.~Mannel, Y.~-M.~Wang,
  [arXiv:1111.1849 [hep-ph]].

\bibitem{Hiller:2001zj}
  G.~Hiller, A.~Kagan,
  Phys.\ Rev.\  {\bf D65}, 074038 (2002)
  [hep-ph/0108074].

\bibitem{hep-ph/0702191} 
  G.~Hiller, M.~Knecht, F.~Legger and T.~Schietinger,
  Phys.\ Lett.\ B\ {\bf 649}, 152  (2007)
  [hep-ph/0702191].

\bibitem{HUTP-90-A071} 
  M.~J.~Dugan and B.~Grinstein,
  Phys.\ Lett.\ B\ {\bf 255}, 583  (1991).


\bibitem{hep-ph/9812358} 
  J.~Charles, A.~Le Yaouanc, L.~Oliver, O.~Pene and J.~C.~Raynal,
  Phys.\ Rev.\ D\ {\bf 60}, 014001  (1999)
  [hep-ph/9812358].


\bibitem{Bauer:2000yr}
  C.~W.~Bauer, S.~Fleming, D.~Pirjol, I.~W.~Stewart,
  Phys.\ Rev.\  {\bf D63}, 114020 (2001)
  [hep-ph/0011336].

\bibitem{Bauer:2001cu}
  C.~W.~Bauer, D.~Pirjol, I.~W.~Stewart,
  Phys.\ Rev.\ Lett.\  {\bf 87}, 201806 (2001)
  [hep-ph/0107002].


\bibitem{hep-ph/0109045} 
  C.~W.~Bauer, D.~Pirjol and I.~W.~Stewart,
  Phys.\ Rev.\ D\ {\bf 65}, 054022  (2002)
  [hep-ph/0109045].

\bibitem{Beneke:2002ph}
  M.~Beneke, A.~P.~Chapovsky, M.~Diehl, T.~Feldmann,
  Nucl.\ Phys.\  {\bf B643}, 431-476 (2002)
  [arXiv:hep-ph/0206152 [hep-ph]].

\bibitem{hep-ph/0211069} 
  C.~W.~Bauer, D.~Pirjol and I.~W.~Stewart,
  Phys.\ Rev.\ D\ {\bf 67}, 071502  (2003)
  [hep-ph/0211069].




\bibitem{hep-ph/0211018} 
  R.~J.~Hill and M.~Neubert,
  Nucl.\ Phys.\ B\ {\bf 657}, 229  (2003)
  [hep-ph/0211018].


\bibitem{hep-ph/0508250} 
  M.~Beneke and D.~Yang,
  Nucl.\ Phys.\ B\ {\bf 736}, 34  (2006)
  [hep-ph/0508250].


\bibitem{hep-ph/0008255} 
  M.~Beneke and T.~Feldmann,
  Nucl.\ Phys.\ B\ {\bf 592}, 3  (2001)
  [hep-ph/0008255].

\bibitem{Beneke:2003pa}
  M.~Beneke, T.~Feldmann,
  Nucl.\ Phys.\  {\bf B685}, 249-296 (2004)
  [hep-ph/0311335].



\bibitem{hep-ph/0402241} 
  M.~Beneke, Y.~Kiyo and D.~s.~Yang,
  Nucl.\ Phys.\ B\ {\bf 692}, 232  (2004)
  [hep-ph/0402241].

\bibitem{hep-ph/0408344} 
  T.~Becher and R.~J.~Hill,
  JHEP\ {\bf 0410}, 055  (2004)
  [hep-ph/0408344].










\bibitem{arXiv:0804.2424} 
  P.~Ball, V.~M.~Braun and E.~Gardi,
  Phys.\ Lett.\ B\ {\bf 665}, 197  (2008)
  [arXiv:0804.2424 [hep-ph]].

\bibitem{arXiv:0811.1812} 
  Y.~-L.~Liu and M.~-Q.~Huang,
  Nucl.\ Phys.\ A\ {\bf 821}, 80  (2009)
  [arXiv:0811.1812 [hep-ph]].


 

\bibitem{Groote:1997yr} 
  S.~Groote, J.~G.~Korner and O.~I.~Yakovlev,
  Phys.\ Rev.\ D {\bf 56}, 3943 (1997)
  [hep-ph/9705447].
\end{thebibliography}
\end{document}